\documentclass[english,superscriptaddress,twocolumn,PRE]{revtex4-1}
\usepackage[latin9]{inputenc}
\usepackage{babel}
\usepackage{docs}
\usepackage{bm}
\usepackage[colorlinks=true,linkcolor=blue]{hyperref}
\usepackage{color}
\usepackage[usenames,dvipsnames]{xcolor}
\usepackage{graphicx}
\usepackage{dcolumn}
\usepackage{natbib}
\usepackage{subfigure}
\usepackage{relsize}
\usepackage{amssymb,amsmath}
\usepackage{epstopdf}
\usepackage{wrapfig}
\usepackage{float}
\usepackage{yfonts}
\usepackage{todonotes}
\usepackage{setspace}
\usepackage{latexsym}
\usepackage{graphicx,color}
\newcommand{\trp}[1]{{{#1}}}


\begin{document}

\title{Minimal model for transient swimming in a liquid crystal}
%\title{\trp{Minimal model for transient swimming in a liquid crystal}}
%%\subtitle{Do you have a subtitle?\\ If so, write it here}
%\author{Madison S. Krieger\inst{1}, Marcelo A. Dias\inst{2} \inst{3}  \and Thomas R. Powers\inst{1} \inst{4}
%% \thanks is optional - remove next line if not needed
%%\thanks{\emph{Present address:} Insert the address here if needed}%
%}                     % Do not remove
%%
%\offprints{thomas\_powers@brown.edu}          % Insert a name or remove this line
%%
%\institute{School of Engineering, Brown University, Providence, RI 02912, USA \and Aalto Science Institute (AscI), Aalto University, Otaniementie 17, FI-02150 Espoo, Finland \and NORDITA, Roslagstullsbacken 23, 106 92 Stockholm, Sweden \ \and Department of Physics, Brown University, Providence, RI 02192, USA }
%
\author{Madison S. Krieger}
\affiliation{School of Engineering, Brown University, Providence, RI 02912, USA}
\author{Marcelo A. Dias}
\affiliation{Aalto Science Institute (AScI), Aalto University, Otaniementie 17, FI-02150 Espoo, Finland} 
\affiliation{NORDITA, Roslagstullsbacken 23, 106 92 Stockholm, Sweden}
\author{Thomas R. Powers}
\affiliation{School of Engineering, Brown University, Providence, RI 02912, USA}
\affiliation{Department of Physics, Brown University, Providence, RI 02192, USA}

\date{Submitted June 4, 2015}%Received: date / Revised version: date}
% abstract is 200 words
\begin{abstract}
When a microorganism begins swimming from rest in a Newtonian fluid such as water, it rapidly attains its steady-state swimming speed since changes in the velocity field spread quickly when the Reynolds number is small.  However, swimming microorganisms are commonly found or studied in complex fluids. Because these fluids have long relaxation times, the time to attain the steady-state swimming speed can also be long. In this article we study the swimming startup problem in the simplest liquid crystalline fluid: a two-dimensional hexatic liquid crystal film. We study the dependence of startup time on anchoring strength and Ericksen number, which is the ratio of viscous to elastic stresses. For strong anchoring, the fluid flow starts up immediately but the liquid crystal field and swimming velocity attain their sinusoidal steady-state values after a time proportional to the relaxation time of the liquid crystal. When the Ericksen number is high, the behavior is the same as in the strong anchoring case for any anchoring strength. We also find that the startup time increases with the ratio of the rotational viscosity to the shear viscosity, and then ultimately saturates once the rotational viscosity is much greater than the shear viscosity.
\end{abstract}
%
%\PACS{
%      {PACS-key}{describing text of that key}   \and
%      {PACS-key}{describing text of that key}
%     } % end of PACS codes
%end of abstract
%
%\authorrunning{\trp{Krieger, Dias, and Powers}}
%\titlerunning{\trp{Transient swimming in a liquid crystal}}
\maketitle

\section{Introduction}
\label{intro}

The hydrodynamics of natural and artificial  microscopic swimmers in Newtonian and complex fluids continues to be an active area of research~\cite{LaugaPowers2009,GagnonKeimArratia2014,MolaeiBarryStockerSheng2014,Li_etal2014,Williams_etal2014}. Many of these studies focus on steady-state swimming. However, natural swimmers start, stop, and change direction, and artificial swimmers must do the same to be of use. Therefore, it is of interest to study transient swimming problems, such as the acceleration of a swimmer from rest. In a Newtonian fluid of viscosity $\mu$ and density $\rho$, the relevant time scale $t_\mathrm{v}$ (`v' for `viscous') for the startup of flow is the time $t_\mathrm{v}=\rho/(\mu q^2)$  for changes in velocity to spread over a distance of order $1/q$. For water and a length scale $1/q\approx1$\,$\mu$m, we have $t_\mathrm{v}\approx 1$\,$\mu$s, which is much shorter than the characteristic beat or rotation frequency of swimming microorganisms, such as $0.001$\,s for \textit{Vibrio alginolyticus}~\cite{Magariyama_etal1995}, $0.01$\,s for \textit{Escherichia coli}~\cite{berg03}, 0.02\,s for \textit{Chlamydomonas reinhardtii}~\cite{KamiyaHasegawa1987} and sea urchin sperm~\cite{GibbonsGibbons1972}, 0.05\,s for human sperm~\cite{Smith_etal2009}, and 0.5\,s for \textit{Caenorhabditis elegans}~\cite{ShenArratia2011}. Calculations for how the swimming speed rapidly rises to its steady state value for an idealized swimmer in a Newtonian fluid were presented by Pak and Lauga~\cite{PakLauga2010}. Complex fluids have additional time scales, which are much longer than $t_\mathrm{v}$, and can be comparable to or even longer than the characteristic beat period. For example, polymer solutions in which swimming has been studied can have relaxation times of the order of seconds~\cite{ShenArratia2011,LiuPowersBreuer2011}. Likewise, the airway mucus encountered by beating cilia has relaxation times of the order of tens of seconds~\cite{GilboaSilberberg1976,LaiWangWirtzHanes2010}.  The startup problem for a swimmer in a viscoelastic fluid has been examined by Elfring and Lauga~\cite{ElfringLauga2014}. Swimming bacteria have also been recently studied
in liquid crystal solutions~\cite{Zhou_etal2013,MushenheimEtAl2013}. For a liquid crystal with Franck elastic constant $K\approx10$\,pN, and shear viscosity $\mu\approx10$\,Pa-s~\cite{Zhou_etal2013}, the characteristic relaxation time $t_\mathrm{e}$ (`e' for `elastic') for distortions of the liquid crystal with length scale $1/q\approx1$\,$\mu$m is $t_\mathrm{e}=\mu/(Kq^2)\approx10$\,s. The long relaxation times of these complex fluids can lead to much longer-lived transient swimming flows than in the Newtonian case. 

In this article we study the startup problem for an idealized swimmer with small amplitude waves in a hexatic liquid crystal film. \trp{The hexatic phase is studied because its theory is mathematically simpler than that of the nematic phase, yet it retains some of the same distinctive features such as Franck elasticity, rotational viscosity, and anchoring effects.} We show that the time required to attain the steady swimming speed is proportional to the relaxation time of the liquid crystal. Since this time is much longer than the viscous startup time $t_\mathrm{v}$, the swimmer first attains the Newtonian swimming speed in a time  comparable to $t_\mathrm{v}$, and then reaches the final swimming speed over a longer time $t_\mathrm{e}$. We study how the evolution of the swimming velocity depends on the anchoring strength of the the liquid crystal at the surface of the swimmer, and also the Ericksen number Er, which is the ratio of viscous to elastic stresses. \trp{For strong anchoring, we find that the swimming speed reaches its ultimate value after a time proportional to the Ericksen number. At high Ericksen number, the behavior of the flow field, the liquid crystal configuration, and the swimming speed is independent of anchoring strength, and is given by the strong anchoring case. When the anchoring strength vanishes, we find that the swimming speed can oscillate as it approaches its final value, and that the swimmer can even reverse direction several times before reaching its steady speed.}

The outline of the paper is as follows. Section~\ref{sec:1} describes the governing equations and timescales. Then \trp{the case of strong anchoring is solved analytically in Sec.~\ref{sec:6}. Section~\ref{arbanch} considers the case of arbitrary anchoring strength in the limits of high and low Ericksen number, paying special attention to the striking case of zero anchoring strength. The conclusion is Sec.~\ref{conclude}.}

%flow and liquid crystal field  is found to first order in the amplitude in Sec.~\ref{sec:4}. Section~\ref{sec:5} contains the solution to the startup problem to second order in amplitude for several cases. First we find the evolution of the swimming speed analytically for the case of strong anchoring. Then we show that when the Ericksen number is high, the evolution of the swimming speed is the same as the strong anchoring case for any anchoring. For low Ericksen number we use a combination of analytic methods and numerical inversion of the Laplace transform to find the swimming speed. Finally, we compute the evolution of the swimming speed for any Ericksen number in the limit of vanishing anchoring strength. We conclude in Sec.~\ref{conclude}.  
%

\section{Setup and governing equations}
\label{sec:1}

\begin{figure}[t]
\includegraphics[width=3.3in]{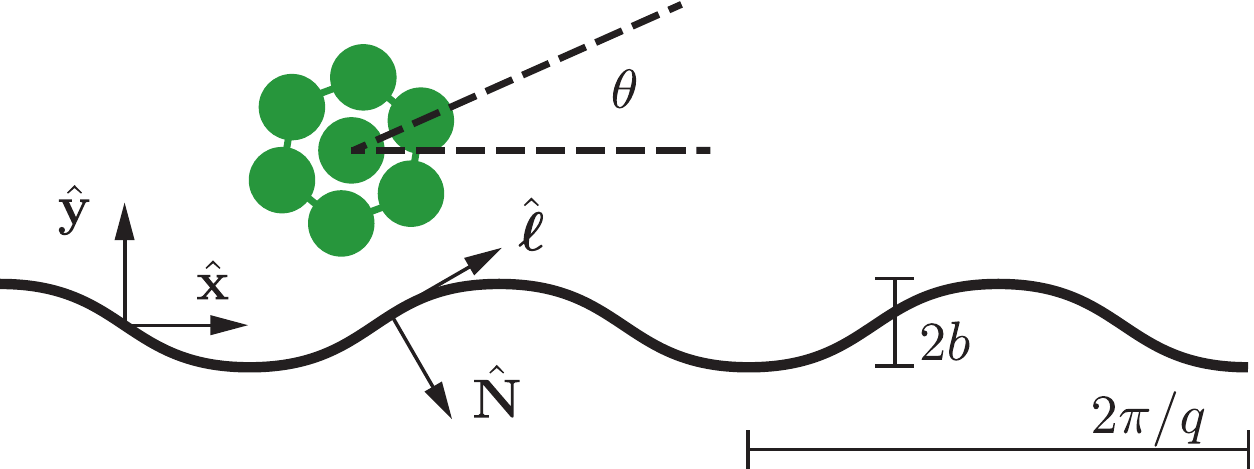}
\caption{(Color online) Sketch of the one-dimensional swimmer and definition of the angle field $\theta$. The liquid crystal particles are not drawn to scale. The outward pointing unit normal vector on the swimmer is $\hat{\mathbf N}$, and the tangent vector to the swimmer is $\hat{\boldsymbol \ell}$.}
\label{setup}
\end{figure}

\subsection{The swimmer}
The swimmer is a line with a prescribed traveling wave. In the frame of the swimmer, the material points on the swimmer have coordinates 
\begin{eqnarray}
x_\mathrm{m}&=&x\\ 
y_\mathrm{m}&=&b\sin[q x-\omega(t)t],
\end{eqnarray}
where $b$ is the amplitude of the wave, $q$ is the wavenumber, and $\omega(t)$ is the time-dependent frequency (Fig.~\ref{setup}).  Following Taylor~\cite{taylor1951}, we assume the amplitude is small compared to the wavelength, $b^2q^2\ll1$. Initially, the frequency vanishes, $\omega(0)=0$, and the fluid is motionless. The frequency rises to its final value $\omega_\infty$ with a characteristic time scale $\trp{t_\mathrm{s}}$ \trp{(`s' for swimmer)}. This characteristic time is long compared to the viscous startup time $t_\mathrm{v}$, but seems to be comparable to or longer than the beat period $2\pi/\omega_\infty$, depending on the species. For example, the switching time from backward to forward swimming, or vice versa, in \textit{V. alginolyticus} is of the order of tenths of seconds~\cite{Xie_etal2011}, about ten times longer than the beat period. When sea urchin spermatozoa are rendered immotile by lowering the pH, the waveform takes a few periods to restart once the pH is raised again~\cite{Goldstein1979}.

\subsection{Hexatic Liquid Crystals}
\label{sec:2}
 The swimmer lies in a two-dimensional hexatic liquid crystal. Recall that a hexatic liquid crystal has six-fold bond-orientational order, described by an angle field $\theta$, where $\theta$ and $\theta+2\pi/6$ denote the same physical configuration~\cite{deGennesProst} (Fig.~\ref{setup}).  We use the notation and conventions of reference~\cite{KriegerSpagnoliePowers2014}, and refer the reader there for more details on the derivation of the  governing equations. We consider incompressible flow, ${\boldsymbol\nabla}\cdot{\mathbf v}=0$, where ${\mathbf v}$ is the velocity. The dynamical equations for velocity and angle field are
\begin{eqnarray}
-{\boldsymbol \nabla}p+\mu\nabla^2{\mathbf v}-K({\boldsymbol\nabla}\theta)\nabla^2\theta+\frac{K}{2}{\boldsymbol\nabla}\times\left(\hat{\mathbf z}\nabla^2\theta\right)={\mathbf 0}\label{fbal}\\
\partial_t \theta+\mathbf{v}\cdot{\boldsymbol\nabla}\theta-\frac{1}{2}\hat{\mathbf z}\cdot{\boldsymbol \nabla}\times\mathbf{v}=\frac{K}{\gamma}\nabla^2\theta,\label{tbal}
\end{eqnarray}
where $\mu$ is shear viscosity, $K$ is the Franck elastic constant, $\hat{\mathbf z}$ is the unit vector perpendicular to the plane of the liquid crystal film, and $\gamma$ is the rotational viscosity~\cite{landau_lifshitz_elas}.
Note that the pressure $p$ has been defined so that it vanishes in equilibrium~\cite{KriegerSpagnoliePowers2014}; in other words, $p$ is the actual pressure minus the equilibrium pressure arising from Franck elasticity~\cite{deGennesProst}. Note also that we have assumed the Reynolds number to be vanishingly small. This approximation is valid because the liquid crystal solutions which motivate our study are 10000 times as viscous as water~\cite{Zhou_etal2013}, and also because we do not attempt to resolve the dynamics on the scale $t_\mathrm{v}$. 

At  the swimmer, we assume no-slip boundary conditions on the velocity field: 
\begin{equation}
\mathbf{v}(x_\mathrm{m},y_\mathrm{m},t)=\dot{y}_\mathrm{m}\hat{\mathbf y}.
\label{noslip}
\end{equation}
Far from the swimmer, the velocity field has an unknown uniform value $\mathbf{v}(y\rightarrow\infty)=U\hat{\mathbf{x}}$ that we must solve for. The flow $U$ in the rest frame is the swimming velocity in the lab frame, which has opposite sign to $U$. We also impose anchoring conditions on the liquid crystal near the swimmer surface. Since we assume the swimmer has a small slope, we may assume the angle $\phi$ between the swimmer tangent vector $\hat{\boldsymbol\ell}$ and the $x$-axis is small. With accuracy of order $b^2q^2$, the anchoring condition becomes~\cite{KriegerSpagnoliePowers2014}
\begin{equation}
K\hat{\mathbf{N}}\cdot{\boldsymbol \nabla}\theta+W(\theta-\phi)=0,\label{anchBC}
\end{equation}
where $\hat{\mathbf{N}}$ is the outward-pointing normal vector \trp{and $W$ is an energy per unit length giving the strength of the anchoring potential.}

As mentioned earlier, the initial condition for the fluid is $\mathbf{v}=\mathbf{0}$. The initial condition for the angle field is the equilibrium configuration, which satisfies $\nabla^2\theta=0$ and the anchoring condition Eq.~(\ref{anchBC}).

\subsection{Time scales, nondimensionalization, and the impulsive startup problem}
\label{sec:3}

There are four time scales in our problem, with a strong separation of the viscous time scale from the others:
\begin{equation}
t_\mathrm{v}\ll \omega_\infty^{-1}<\trp{t_\mathrm{s}}<t_\mathrm{e}.
\end{equation}

To make the governing equations dimensionless, we use $1/\omega_\infty$ as the units of time and $1/q$ as the units of length.  Then, with $p$ measured in units of $\mu\omega_\infty$, we have~\cite{KriegerSpagnoliePowers2014}
\begin{eqnarray}
-\boldsymbol{\nabla} p+\nabla^2\mathbf{v}-\frac{1}{\mathrm{Er}}\boldsymbol{\nabla}\theta(\nabla^2\theta)+\frac{1}{2\mathrm{Er}}\boldsymbol{\nabla}\nabla^2\theta\times\hat{\mathbf{z}}&=&0\label{force}\\
\partial_t\theta+\mathbf{v}\cdot\boldsymbol{\nabla}\theta-\frac{1}{2}\hat{\mathbf{z}}\cdot\boldsymbol{\nabla}\times\mathbf{v}-\frac{\mu}{\gamma}\frac{1}{\mathrm{Er}}\nabla^2\theta&=&0,\label{torque}
\end{eqnarray}
where 
\begin{equation}
\mathrm{Er}=\frac{\mu\omega_\infty}{Kq^2}.\label{Erdef}
\end{equation} 
%and $\boldsymbol{\nabla}^3\theta\equiv\boldsymbol{\nabla}\nabla^2\theta$.
It is convenient to use  the stream function $\psi$, which automatically enforces the constraint of incompressibility by the definition $\mathbf{v}=\boldsymbol{\nabla}\times(\psi \hat{\mathbf{z}})$. In terms of $\psi$, the governing equations are
\begin{eqnarray}
\nabla^4\psi+\frac{1}{2\mathrm{Er}}\nabla^4\theta+\frac{1}{\mathrm{Er}}\hat{\mathbf{z}}\cdot\boldsymbol{\nabla}\theta\times\boldsymbol{\nabla}\nabla^2\theta&=&0\label{psieqn1}\\
\partial_t\theta+\frac{1}{2}\nabla^2\psi+\hat{\mathbf{z}}\cdot\boldsymbol{\nabla}\theta\times\boldsymbol{\nabla}\psi-\frac{\mu}{\gamma}\frac{1}{\mathrm{Er}}\nabla^2\theta&=&0.\label{psieqn2}
\end{eqnarray}

In dimensionless form, the boundary condition for the angle field $\theta(x_\mathrm{m},y_\mathrm{m})$ at the swimmer is
\begin{equation}
\hat{\mathbf{N}}\cdot{\boldsymbol \nabla}\theta+w(\theta-\phi)=0,\label{dimlessanchBC}
\end{equation}
where $w=W/(Kq)$. The dimensionless no-slip boundary condition at $(x_\mathrm{m},y_\mathrm{m})$ is
\begin{equation}
\mathbf{v}(x_\mathrm{m},y_\mathrm{m})=-\epsilon \frac{\mathrm{d}[\omega(t)t]}{{\mathrm d}t}\cos\left[x- \omega(t)t\right]\hat{\mathbf y},\label{dimlessvBC}
\end{equation}
where we have introduced $\epsilon=bq$. Note that we are also measuring frequency in units of $\omega_\infty$
To simplify the analysis, we will consider the impulsive startup problem and suppose that $\omega_\infty\tau\ll1$, even though $\tau$ is typically a longer time scale than $1/\omega_\infty$. 
Therefore, we suppose the dimensionless frequency impulsively jumps from zero to unity at $t=0$.  Denoting by $H(t)$ the unit Heaviside step function, we have
\begin{equation}
\mathbf{v}(x_\mathrm{m},y_\mathrm{m})=-\epsilon \frac{\mathrm{d}[H(t)t]}{{\mathrm d}t}\cos\left[x- H(t)t\right]\hat{\mathbf y}.\label{newdimlessvBC}
\end{equation}
Our task is to solve for the swimming speed $-U$ as a function of time. We expand in powers of $\epsilon$, 
\begin{eqnarray}
\mathbf{v}&=&\mathbf{v}^{(1)}+\mathbf{v}^{(2)}+\cdots\\
\mathbf{\theta}&=&\mathbf{\theta}^{(1)}+\mathbf{\theta}^{(2)}+\cdots
\end{eqnarray}
where the superscript denotes the power of $\epsilon$.

\section{Strong anchoring}
\label{sec:6}

First we consider the case of strong anchoring, $w\rightarrow\infty$, since in this limit we can get explicit expressions for $\theta$ and $U$ as functions of time for impulsive startup of the beating. 

%\subsection{First order problem}
In terms of the stream function, the first-order parts of equations (\ref{psieqn1}--\ref{psieqn2}) are
\begin{eqnarray}
\nabla ^4 {\psi}^{(1)} + \frac{1}{2 \mathrm{Er}} \nabla ^4 {\theta}^{(1)} = 0 \label{fbal1}\\
\partial_t{\theta}^{(1)}+\frac{1}{2} \nabla ^2 {\psi}^{(1)}  -\frac{\mu}{\gamma \mathrm{Er}} \nabla ^2 {\theta}^{(1)} &=&0.\label{fbal2}
\end{eqnarray}
The initial conditions are
\begin{eqnarray}
\theta^{(1)}(x,y,t=0)&=&\epsilon\mathrm{e}^{-y}\cos x\\
\psi^{(1)}(x,y,t=0)&=&0,
\end{eqnarray}
and the boundary conditions are
\begin{eqnarray}
\boldsymbol{\nabla}\psi^{(1)}|_{y=0}&=&\trp{-}\epsilon\frac{\mathrm{d}}{\mathrm{d}t}\sin[x-tH(t)]\hat{\mathbf x}\\
\theta^{(1)}|_{y=0}&=&\epsilon\cos(x-t).
\end{eqnarray}
Note that the angle field has some initial distortion  due to the strong-anchoring condition. Furthermore, the initial condition for the angle field has exactly the same spatial form as the sinusoidal-steady state solution for $\theta^{(1)}$ that was found in~\cite{KriegerSpagnoliePowers2014}. Thus, there is no need for distortions in the angle-field to spread once the swimmer waveform starts up; the angle field simply starts to oscillate in time with the same frequency as the swimmer waveform. The fluid velocity, however, changes discontinuously, since we do not resolve dynamics on the small timescale $t_\mathrm{v}$ (studied in~\cite{PakLauga2010}). \trp{For $t>0$, we find} 
\begin{eqnarray}
v_x^{(1)}&=&-\epsilon y\mathrm{e}^{-y}\sin(x-t)\\
v_y^{(1)}&=&-\epsilon (1+y)\mathrm{e}^{-y}\cos(x-t)\\
\theta^{(1)}&=&\epsilon\mathrm{e}^{-y}\cos(x-t).
\end{eqnarray}
These results can also be found by directly solving the equations~(\ref{fbal1}--\ref{fbal2}) using Laplace transforms.

To see how the final sinusoidal steady-state solution emerges, we must turn to the second-order equations:
\begin{eqnarray}
&-&\boldsymbol{\nabla}p^{(2)}+\nabla^2\mathbf{v}^{(2)}\nonumber\\&=&-\frac{1}{\mathrm{Er}}\left[\frac{1}{2}\boldsymbol{\nabla}\times(\hat{\mathbf z}\nabla^2\theta^{(2)})-\boldsymbol{\nabla}\theta^{(1)}\nabla^2\theta^{(1)}\right], \label{force2}
\end{eqnarray}
and
\begin{equation}
\frac{1}{\mathrm{Er}}\frac{\mu}{\gamma}\nabla^2\theta^{(2)}=\partial_t\theta^{(2)}+\mathbf{v}^{(1)}\cdot\boldsymbol{\nabla}\theta^{(1)}-\frac{1}{2}\hat{\mathbf z}\cdot\boldsymbol{\nabla}\times\mathbf{v}^{(2)}. \label{torque2}
\end{equation}
%\frac{1}{\mathrm{Er}}\frac{\mu}{\gamma}\nabla^2\theta^{(2}}=\partial_t\theta^{(2)}+\mathbf{v}^{(1)}\cdot\boldsymbol{\nabla}\theta^{(1)}-\frac{1}{2}\hat{\mathbf z}\cdot\boldsymbol{\nabla}\times\mathbf{v}^{(2)}
%\end{eqnarray}
%We average over a spatial wavelength %and use the fact that the first-order angle field is harmonic 
%to write 
\trp{Averaging over a spatial wavelength, these equations become}
\begin{eqnarray}
\partial_y^2\langle v_x^{(2)}\rangle+\frac{1}{2\mathrm{Er}}\partial_y^3\langle\theta^{(2)}\rangle=f\label{secorde1}\\
\partial_t\langle\theta^{(2)}\rangle=\frac{1}{\mathrm{Er}}\frac{\mu}{\gamma}\partial^2_y\langle\theta^{(2)}\rangle-\frac{1}{2}\langle \partial_y v_x^{(2)}\rangle-g\label{secorde2},
\end{eqnarray}
where \trp{$f=\langle\partial_x\theta^{(1)}\nabla^2\theta^{(1)}\rangle/\mathrm{Er}$ and }$g\equiv\langle\mathbf{v}^{(1)}\cdot\boldsymbol{\nabla}\theta^{(1)}\rangle.$ \trp{For strong anchoring,} 
\begin{eqnarray}
f&=&0\\
g&=&(\epsilon^2/2)(1+2y)\exp(-2y)H(t).
\end{eqnarray}
The initial conditions are $\langle\theta^{(2)}\rangle|_{t=0}=0$ and $\langle\mathbf{v}^{(2)}\rangle|_{t=0}=0$, and the boundary conditions are
\begin{eqnarray}
\langle v_x^{(2)}\rangle|_{y=0}&=&-\langle y_\mathrm{m}\partial_y v_x^{(1)}\rangle|_{y=0}=\epsilon^2/2\label{sabc1}\\
\langle\theta^{(2)}\rangle|_{y=0}&=&0.\label{sabc2}
\end{eqnarray}
\trp{Note that to arrive at~(\ref{sabc1}--\ref{sabc2}), we expanded the boundary condition at $y=y_\mathrm{m}$ and used the fact that the strong-anchoring angle field is out of phase with the wave $y=y_\mathrm{m}$.}

Using $\mathcal{L}$ to denote the Laplace transform,
\begin{equation}
\mathcal{L}\{f(t)\}=\int_0^\infty\exp(-st)f(t)\mathrm{d}t,
\end{equation}
let $\tilde V(s)\equiv\mathcal{L}\langle v_x^{(2)}\rangle$, and $\tilde \Theta(s)\equiv\mathcal{L}\langle\theta^{(2)}\rangle$.
Then
\begin{eqnarray}
\tilde V''+\frac{1}{2\mathrm{Er}}\tilde\Theta'''&=&0\label{leq1}\\
s\tilde\Theta-\frac{1}{\mathrm{Er}}\frac{\mu}{\gamma}\tilde\Theta''+\frac{1}{2}\tilde V'&=&-\tilde g,\label{leq2}
\end{eqnarray}
where $\tilde g=g/s$ and the primes denote differentiation with respect to $y$.
Note that (\ref{leq1}) and the fact that $\theta^{(2)}$ and $v_x^{(2)}$ are finite at $y\rightarrow\infty$ implies that $\tilde V'+\tilde\Theta''/2/\mathrm{Er}$ is a constant. Thus, (\ref{leq2}) becomes diffusion with a source,
\begin{equation}
s\tilde\Theta-D\tilde\Theta''=-\tilde g,
\end{equation}
where the diffusion constant is
\begin{equation}
D=\frac{1}{\mathrm{Er}}\left(\frac{\mu}{\gamma}+\frac{1}{4}\right).
\end{equation}
\begin{figure}[t]
\includegraphics[width=3.35in]{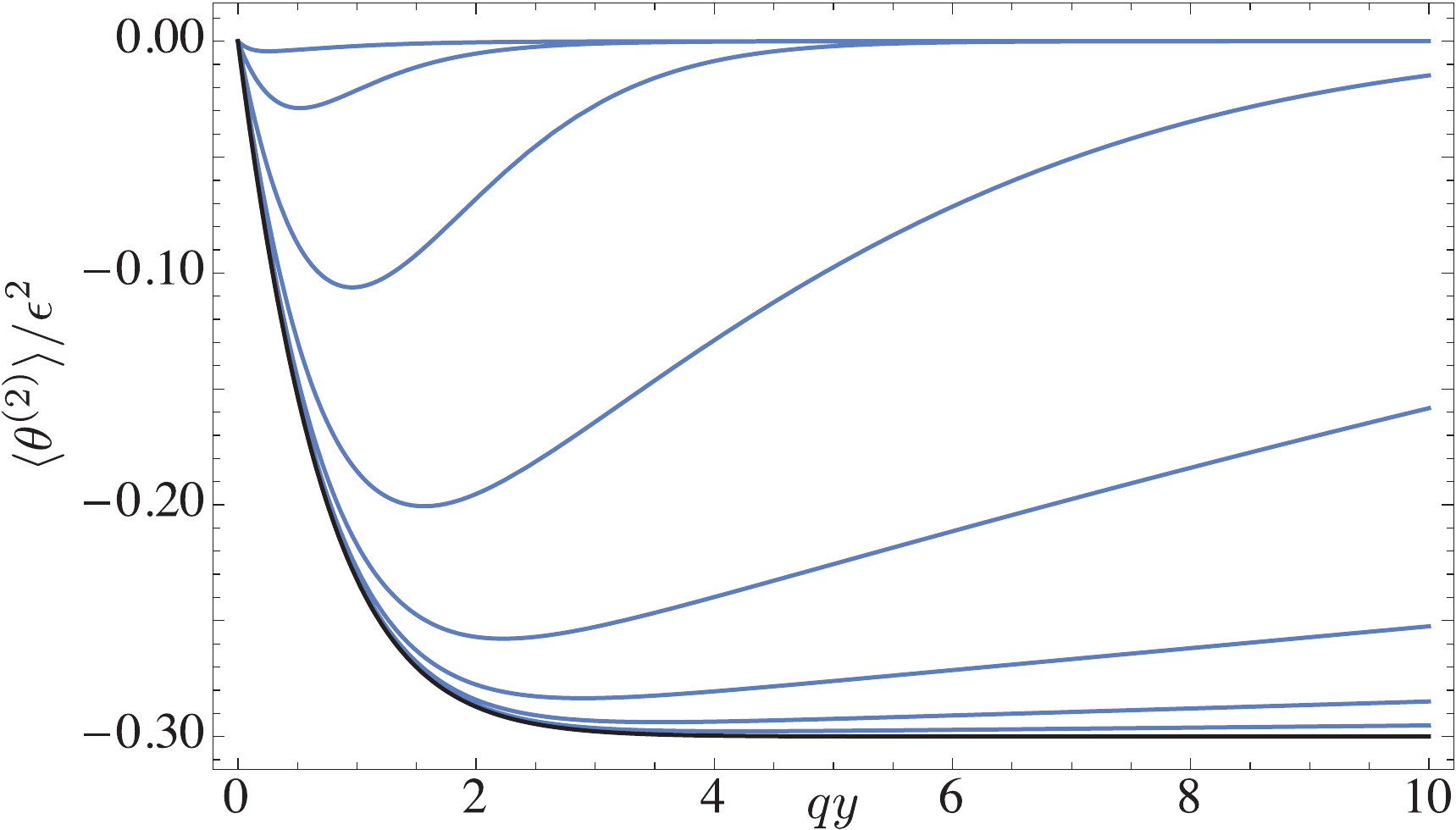}
\caption{(Color online)
Time evolution of the average second-order angle field $\langle\theta^{(2)}\rangle$ for $\mathrm{Er}=1$ and $\mu=\gamma$, or $D=5/4$. Starting from the top of the figure, the first eight curves (blue) correspond to $\omega_\infty t=10^{-2}$, $10^{1}$, ..., $10^{5}$, and the last curve (black) corresponds to the steady-state value, or $\omega_\infty t\rightarrow\infty$.}
\label{thetaspread}
\end{figure}
The solution is 
\begin{eqnarray}
\tilde\Theta&=&\frac{\epsilon^2}{2s}\frac{s-12D}{(s-4D)^2}\mathrm{e}^{-y\sqrt{s/D}}\nonumber\\
&+&\frac{\epsilon^2}{2s}\frac{12D-s+(8D-2s)y}{(s-4D)^2}\mathrm{e}^{-2y}.
\end{eqnarray}
Note that we recover the proper long-time limit of the second-order angle field in steady-state swimming~\cite{KriegerSpagnoliePowers2014}:
\begin{equation}
\lim_{s\rightarrow0}s\tilde\Theta=\epsilon^2\frac{\gamma\mathrm{Er}}{2(4\mu+\gamma)}\left[(3+2y)\mathrm{e}^{-2y}-3\right].
\end{equation}
The angle field $\tilde\Theta$ has an explicit inverse Laplace transform:
\begin{eqnarray}
\langle\theta^{(2)}\rangle&=&\frac{\epsilon^2}{16D}\left\{\left(6+4y\right)\mathrm{e}^{-2y}-6\,\mathrm{erfc}\left[y/(2\sqrt{D t})\right]\right.\nonumber\\
&+&\left(8Dt-3-2y\right)\nonumber\\
&\times&\left.\left[\mathrm{e}^{4Dt-2y}\,\mathrm{erfc}\left(z_-\right)-\mathrm{e}^{2y}\,\mathrm{erfc}\left(z_+\right)\right]\right\},\label{strongtheta}
\end{eqnarray}
where $\mathrm{erfc}$ is the complementary error function~\cite{AbramowitzStegun1964}, and
\begin{equation}
z_\pm\equiv\frac{4Dt\pm y}{2\sqrt{D t}}.
\end{equation}
\trp{Note that since $D\propto1/\mathrm{Er}$, we have $\langle\theta^{(2)}\rangle\propto\mathrm{Er}$.}
Figure~\ref{thetaspread} shows how $\langle\theta^{(2)}\rangle$ evolves to its final steady-state value. Note that for any finite $t$, the second-order angle field vanishes at $y\rightarrow\infty$. However, for a given $y\gtrapprox1$, we can always wait long enough for the angle field to reach its steady-state value of $-3\epsilon^2\gamma \mathrm{Er}/(8\mu+\gamma)$.
Despite the complicated form of Eq.~(\ref{strongtheta}), close examination of \trp{the} plots of $\langle\theta^{(2)}\rangle$ for various $t$ \trp{(Fig.~\ref{thetaspread})} reveals that the $y$-value at which $|\langle\theta^{(2)}\rangle|$ attains half its maximum value increases like $t^{1/2}$, as expected for diffusion. 
%\color{red}Discuss\color{black} 

%\color{red}
%We see that this sourced diffusion causes the mean squared change in the angle field to propagate faster than unsourced diffusion, with $(\langle \Theta \rangle)^2 \sim t^{3/2}$
%\color{black} 

We can also solve for the velocity field in Laplace space:
%, which has a  more complicated expression:
\begin{equation}
\tilde V=v_1\mathrm{e}^{-y\sqrt{s/D}}+v_2\mathrm{e}^{-2y}+\tilde U, \label{strongV}
\end{equation}
where
\begin{eqnarray}
v_1&=&\epsilon^2\frac{\sqrt{{s^3/D}}-12\sqrt{sD}}{4\mathrm{Er}s(s-4D)^2}\\
v_2&=&\epsilon^2\frac{-4sy+16D(1+y)}{4\mathrm{Er}s(s-4D)^2},
\end{eqnarray}
%Taking the limit $y\rightarrow\infty$ reveals the swimming speed in the frequency domain,
and the term $\tilde U$ remaining in the limit $y\rightarrow\infty$ is the swimming speed in the frequency domain,
\begin{equation}
\tilde U=\left[\frac{1}{2s}-\frac{4+\sqrt{s/D}}{s\mathrm{Er}(2\sqrt{D}+\sqrt{s})^2}\right]\epsilon^2.
\end{equation}
Note that the long-time limit of the swimming speed agrees with the steady-state result~\cite{KriegerSpagnoliePowers2014}:
\begin{equation}
\lim_{s\rightarrow0} s\tilde U=U_\infty=\frac{\epsilon^2}{2}\frac{4\mu-\gamma}{4\mu+\gamma}.
\end{equation}
By inspection of $\tilde\Theta$, $\tilde V$, and $\tilde U$, we observe that the characteristic time scale for disturbances to diffuse over a distance $L$ is $L/\sqrt{D}\propto L\sqrt{\mathrm{Er}\gamma/(\mu+\gamma/4)}$. 

%Our expression for $\tilde U$ has an analytic inverse Laplace transform which is too cumbersome to write. However we can expand $\tilde U$ for large $s$,
%\begin{equation}
%\tilde U\sim\epsilon^2\left[\frac{1}{2s}-\frac{1}{4\mathrm{Er}\sqrt{s^3 D}}+\frac{\sqrt{D}}{\mathrm{Er}s^{5/2}}+\mathcal{O}\left(\frac{1}{s^3}\right)\right]
%\end{equation}
%and use Watson's lemma~\cite{Davies2002} and take the inverse Laplace transform term-by-term to find
%\begin{equation}
%U\sim\frac{\epsilon^2}{2}\left[1-\sqrt{\frac{t}{\pi D}}\frac{1}{\mathrm{Er}}+\frac{8}{3}\sqrt{\frac{D t^3}{\pi}}\frac{1}{\mathrm{Er}}+\mathcal{O}\left(t^2\right)\right].\label{smallt}
%\end{equation}

%For large $t$ I am still uncertain about how to proceed. We can expand $\tilde U$ for small $s$ to find
%\begin{equation}
%\tilde U\sim\epsilon^2\left[\frac{U_\infty}{s}+\frac{3}{4}\frac{1}{\mathrm{Er}\sqrt{s D^3}}-\frac{1}{2\mathrm{Er}D^2}+\mathcal{O}\left(\sqrt{s}\right)\right]
%\end{equation}
%If we only had the $1/\sqrt{s}$ term in addition to the steady-state $1/s$ term, then we might say the speed approaches $U_\infty$ with a $1/\sqrt{t}$ decay, i.e. 
%\begin{equation}
%U\sim\epsilon^2\left[U_\infty+\frac{3}{4}\frac{1}{\mathrm{Er}\sqrt{\pi t D^3}}+\dots\right].
%\end{equation} 
%But the other terms [eg $\mathcal{O}(\sqrt{s})$] seem to cause problems since they do not have inverse Laplace transforms....

\begin{figure}[t]
\includegraphics[width=3.3in]{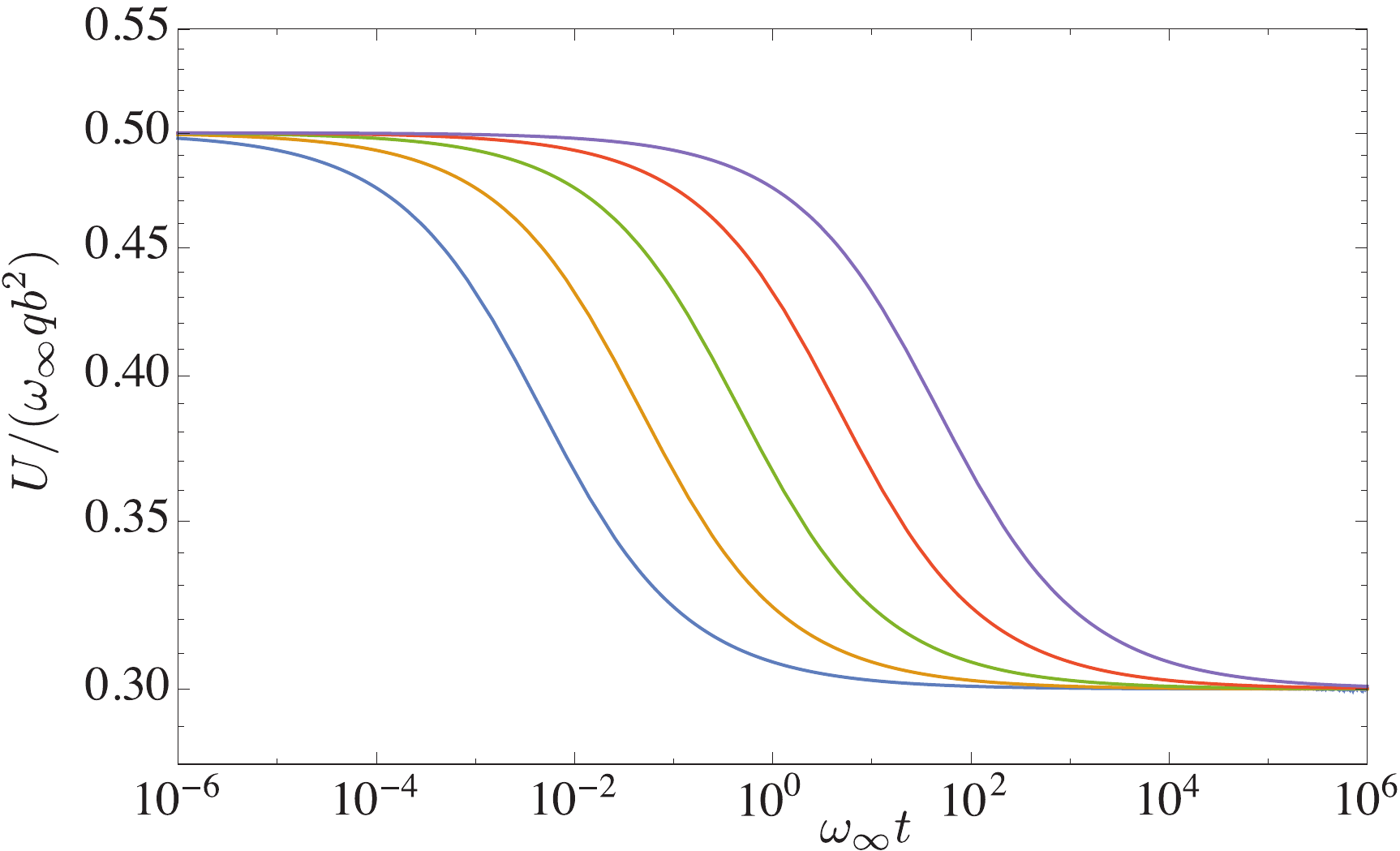}
\caption{(Color online)
Dimensionless \trp{swimming} speed vs dimensionless time on a log-log scale for $\mu=\gamma$ and Er ranging from (from left to right) Er$=0.01$ (blue), Er=$0.1$ (gold), Er=1 (green), Er$=10$ (red), and Er$=100$ (purple).}
\label{Uvt}
\end{figure}
We were unable to find an analytic form for the velocity field in the time domain. However, we can solve for the swimming speed in the time domain by integrating Eq.~(\ref{leq1}) in the time domain with respect to $y$. Let $V=\langle v_x^{(2)}\rangle$ and $\Theta=\langle\theta^{(2)}\rangle$. One integration yields $V'+\Theta''/(2\mathrm{Er})=0$, since all $y$-derivatives of $V$ and $\Theta$ must vanish at $y\rightarrow\infty$. One more integration shows that $V+\Theta'/(2\mathrm{Er})$ is constant; this constant must equal the flow speed at $y\rightarrow\infty$, again because $\Theta'$ vanishes at $y\rightarrow \infty$. Thus,
\begin{equation}
U=V(y,t)+\frac{1}{2\mathrm{Er}}\frac{\partial \Theta}{\partial y}\left(y,t\right).
\end{equation}
But $V(0,t)=\epsilon^2/2$; therefore
\begin{equation}
U=\frac{\epsilon^2}{2}+\left.\frac{1}{2\mathrm{Er}}\partial_y\Theta\right|_{y=0}\label{Ueqn}.
\end{equation}
Substituting Eq.~(\ref{strongtheta}) into (\ref{Ueqn}) yields
\begin{equation}
U=U_\infty+U_\mathrm{tr},
\end{equation}
where
\begin{equation}
U_\infty=\frac{1}{2}-\frac{1}{4D\mathrm{Er}}=\frac{1}{2}\frac{4\mu-\gamma}{4\mu+\gamma}
\end{equation}
is the steady-state speed at $t\rightarrow\infty$~\cite{KriegerSpagnoliePowers2014}, and 
\begin{equation}
U_\mathrm{tr}=\epsilon^2\left[\frac{1}{2\mathrm{Er}}\sqrt{\frac{t}{D\pi}}
+\frac{1}{\mathrm{Er}}\left(\frac{1}{4D}-t\right)\mathrm{e}^{4Dt}\,\mathrm{erfc}\left(2\sqrt{Dt}\right)\right] \label{strongU}
\end{equation}
is a transient contribution that decays for large $t$ as
\begin{equation}
U_\mathrm{tr}\sim\frac{3}{16\left(\mu/\gamma+1/4\right)^{3/2}}\sqrt{\frac{\mathrm{Er}}{\pi t}}+\mathcal{O}(1/t^{3/2}).\label{Utr}
\end{equation}
The evolution of the swimming speed in time for various Er is shown in Fig.~\ref{Uvt}. Since the curves in Fig.~\ref{Uvt} are equally spaced in $\log(\mathrm{Er})$ and are also equally spaced on the log-log plot, the time to reach the midway point between $\omega_\infty qb^2/2$ and $U_\infty$ increases linearly with Er. We can also see that the time to reach the steady state increases linearly with $\mathrm{Er}$ from Eq.~(\ref{Utr}): the \trp{dimensionless} time \trp{$t^*$} it takes $U_\mathrm{tr}$ to decay to a small fraction of the steady-state speed is
% proportional to Er, or equivalently, the relaxation time $\mu/(Kq^2)$ of the liquid crystal.
\begin{equation}
\trp{t^*\propto\frac{\mathrm{Er}}{(\mu/\gamma+1/4)^3}.}
\end{equation}
\trp{Note that $t^*\propto\mathrm{Er}(\gamma/\mu)^3$ for $\gamma\ll\mu$, and $t^*\propto\mathrm{Er}$ for $\gamma\gg\mu$.}

%For small $t$ we recover the expansion~(\ref{smallt}) (this time in terms of Er)
%\begin{equation}
%U\sim\epsilon^2\left[\frac{1}{2}-\sqrt{\frac{t}{\pi\mathrm{Er}\left(1+4\mu/\gamma\right)}}+\frac{2}{3}\sqrt{\frac{t^3\left(1+4\mu/\gamma\right)}{\pi\mathrm{Er}^3}}+\cdots\right],
%\end{equation}
%and for large $t$ we find
%\begin{equation}
%U\sim U_\infty+\frac{3\epsilon^2}{2}\sqrt{\frac{1}{\pi  \mathrm{Er}\,t\left(1+4\mu/\gamma\right)^3}}
%-\frac{5\epsilon^2}{4}\sqrt{\frac{\mathrm{Er}^3}{\pi t^3\left(1+4\mu/\gamma\right)^5}}\cdots
%\end{equation}

%\color{red} Note that I cut all the small-time and long-time asymptotics, it seems redundant when we have the full solution. \color{black}

\section{Arbitrary anchoring strength}
\label{arbanch}
\trp{We now turn to the swimming problem with finite but arbitrary anchoring strength $w$. Again we work in powers of the dimensionless amplitude $\epsilon$. To first order we can find an analytic form for the Laplace transforms of the angle field and stream function. At second order we consider limits such as large Ericksen number, small Ericksen number, and vanishing anchoring strength.} 

\subsection{First-order swimming problem}
\label{sec:4}

%We can write \trp{the governing equations}%Eq.
%~(\ref{force}\trp{--}\ref{torque}) to first order in $\epsilon$ in Laplace space:
\trp{To write the Laplace transform of the governing equations}~(\ref{force}\trp{--}\ref{torque}) to first order in $\epsilon$, \trp{we need to find the initial value of the angle field, since $\mathcal{L}\{\partial_t\theta^{(1)}\}=-\theta^{(1)}(t=0)+s\mathcal{L}\{\theta^{(1)}\}$. The initial condition for the angle field is that it is the equilibrium field obeying the anchoring condition~(\ref{anchBC}) for the initial swimmer shape, $y_\mathrm{m}=-\mathrm{i}\exp\mathrm{i}x$ in complex notation, which leads to} 
\begin{eqnarray}
0&=&\nabla ^4 \tilde{\psi}^{(1)} + \frac{1}{2 \mathrm{Er}} \nabla ^4 \tilde{\theta}^{(1)} \label{laplin1}\\
s \tilde{\theta}^{(1)} &=& \frac{\mu}{\gamma \mathrm{Er}} \nabla ^2 \tilde{\theta}^{(1)} - \frac{1}{2} \nabla ^2 \tilde{\psi}^{(1)} + \frac{w \epsilon}{1+w} e^{\mathrm{i} x-y}\trp{.}\label{laplin2}
\end{eqnarray}
The solutions 
\trp{to Eqns.~(\ref{laplin1}--\ref{laplin2})} are 
%given by the real parts of 
\begin{eqnarray}
\tilde{\psi}^{(1)}&=&(c_0+c_1 y)\mathrm{e}^{-y+\mathrm{i}x}-\frac{1}{2\mathrm{Er}} \tilde\theta^{(1)}\label{psieqn}\\
\tilde{\theta}^{(1)}&=&\left[\frac{1}{s}\left(c_1+\frac{w \epsilon}{1+w}\right)\mathrm{e}^{-y}+c_2 \mathrm{e}^{k y}\right]\mathrm{e}^{\mathrm{i}x},\label{thetaeqn}
\label{foeqns}
\end{eqnarray}
where 
\begin{equation}
k=-\sqrt{1+\frac{4s \gamma  \mathrm{Er}}{\gamma+ 4\mu }},\label{keqn}
\end{equation} 
and the %constants 
\trp{$s$-dependent coefficients} $c_0$, $c_1$, and $c_2$ are determined by 
the no-slip and anchoring boundary conditions at the surface of the swimmer. In Laplace space, the  no-slip boundary condition $\mathbf{v}(x_\mathrm{\trp{m}},y_\mathrm{\trp{m}})=(x_\mathrm{\trp{m}},\partial_t y_\mathrm{\trp{m}})$ %is
\trp{becomes}
\begin{equation}
\left.\left(\partial_y\tilde{\psi}^{(1)},-\partial_x\tilde{\psi}^{(1)}\right)\right|_{y=0}=\left(0,-\epsilon\frac{e^{\mathrm{i} x}}{\mathrm{i}+s}\right)\label{bc1}
\end{equation}
to first order in dimensionless form. 
Likewise, to first order, the anchoring boundary condition (\ref{dimlessanchBC}) %is
\trp{becomes}
\begin{equation}
-\left.\partial_y\tilde{\theta}^{(1)}\right|_{y=0}+w\left(\left.\tilde{\theta}^{(1)}\right|_{y=0}-\epsilon\frac{e^{\mathrm{i} x}}{\mathrm{i}+s}\right)=0.\label{bc1anch}
\end{equation}
\trp{Although analytic expressions for $c_0$, $c_1$, and $c_2$ and also $\tilde{\psi}^{(1)}$ and $\tilde{\theta}^{(1)}$ can be computed for arbitrary $w$, their lengthy form prevents us from displaying them here. In the next two subsections we consider the %asymptotic 
limits of large and small Ericksen number. Just as in the strong anchoring case and the sinusoidal steady-state case~\cite{KriegerSpagnoliePowers2014}, there is no swimming speed to first order.}

%Analytic expressions for the constants $c_0$, $c_1$, and $c_2$ and the first order quantities $\hat{\psi}^{(1)}$ and $\hat{\theta}^{(1)}$ may be found; however, these expressions are too unwieldy to display here. We consider the limiting values of large and small Er below. It should be noted that since the solutions~(\ref{psieqn}, \ref{thetaeqn}) average to zero over $x$, {and since the solution $\psi^{(1)}=v^{(1)}_0 y$ is ruled out by the no-slip boundary condition at $y=0$},  there is no swimming speed to first order. 

\subsection{Asymptotic solution at high Ericksen number}
\label{sec:9}

\trp{Our analysis in the limit of $\mathrm{Er}\gg1$ is similar to that in the sinusoidal steady-state problem~\cite{KriegerSpagnoliePowers2014}. Expanding the coefficients from the first-order problem to leading order in inverse powers of Ericksen number, we find} 
%When the Ericksen number is large,  the form of the decay rate $k$ [Eqn.~(\ref{keqn})] implies a boundary layer near the swimmer of thickness 
%\begin{equation}
%\delta\propto\sqrt{\frac{\gamma+4\mu}{\gamma}}\frac{1}{\sqrt{s \mathrm{Er}}}
%\end{equation}
%in both the angle field and flow field, as long as $\gamma\neq0$. The strength of the anchoring $w=W/(Kq)$ does not affect the boundary layer thickness. Inside the boundary layer, adjacent to the swimmer, elastic forces and torques balance with viscous forces and torques. Outside the boundary layer, the elastic effects can be disregarded, and the local rate of rotation of the angle field is equal to the  local rate of rotation of the fluid. The boundary layer has a small effect on the power dissipated, swimming speed, and flux. To see why, 
% we expand the exact solutions of the first-order problem in  powers of $1/\sqrt{\mathrm{Er}}$ to find
%
%\begin{eqnarray}
%c0 &=& - \frac{\mathrm{i} \epsilon}{\mathrm{i} + s} + \frac{1}{Er} \frac{\epsilon (s w - \mathrm{i})}{2s (\mathrm(i)+s)(1+w)} + \mathcal{O}(\frac{1}{Er^{3/2}}) \\
%c1 &=&  - \frac{\mathrm{i} \epsilon}{\mathrm{i} + s} - \frac{1}{Er} \frac{\mathrm{i} \epsilon}{2 s (\mathrm{i}+s)} + \mathcal{O}(\frac{1}{Er^{3/2}}) \\
%c2 &=& \frac{1}{\sqrt{Er}}\frac{\mathrm{i} \epsilon \sqrt{1/4 + \mu / \gamma}}{(\mathrm{i}+s)\sqrt{s^3}} - \frac{1}{Er} \frac{\mathrm{i} w \epsilon (1/4 + \mu / \gamma)}{s^2 (\mathrm{i}+s)}  + \mathcal{O}(\frac{1}{Er^{3/2}})
%\end{eqnarray}
\begin{eqnarray}
c_0 &=& - \frac{\mathrm{i} \epsilon}{\mathrm{i} + s} + \mathcal{O}\left(\frac{1}{\mathrm{Er}}\right)\\
% \frac{\epsilon (s w - \mathrm{i})}{2s (\mathrm(i)+s)(1+w)} + \m athcal{O}(\frac{1}{Er^{3/2}}) \\
c_1 &=&  - \frac{\mathrm{i} \epsilon}{\mathrm{i} + s} 
+\mathcal{ O}\left(\frac{1}{\mathrm{Er}}\right) \\
% \frac{\mathrm{i} \epsilon}{2 s (\mathrm{i}+s)} + \mathcal{O}(\frac{1}{Er^{3/2}}) \\
c_2 &=& \frac{1}{\sqrt{\mathrm{Er}}}\frac{\mathrm{i} \epsilon \sqrt{1/4 + \mu / \gamma}}{(\mathrm{i}+s)\sqrt{s^3}}
+\mathcal{O}\left( \frac{1}{\mathrm{Er}} \right)
%\frac{\mathrm{i} w \epsilon (1/4 + \mu / \gamma)}{s^2 (\mathrm{i}+s)}  + \mathcal{O}(\frac{1}{Er^{3/2}})
\end{eqnarray}
\trp{With an accuracy of $\mathcal{O}(1/\sqrt{\mathrm{Er}})$, the coefficients $c_0$, $c_1$, and $c_2$ are independent of the anchoring strength, and take the same values as in the strong-anchoring problem; the rapidly varying component, proportional to $\exp{k y}$, which does not arise in the strong-anchoring limit, has an amplitude of $\mathcal{O}(1/\sqrt{\mathrm{Er}})$. Thus we conclude that the first-order flow field and angle field at large Ericksen number and \textit{arbitrary} anchoring is well-approximated by the solutions to the \textit{strong}-anchoring problem.}

\trp{The second-order flow and angle field at large Ericksen number and arbitrary anchoring strength is also given by the strong-anchoring solutions. To see why, note that large Er, the second-order equations~(\ref{secorde1}--\ref{secorde2}) are identical to the strong-anchoring equations since $f$ and $g$ depend only on first-order quantities. 
For arbitrary $w$, the boundary conditions at second order are 
\begin{equation}
\langle v_x^{(2)}\rangle=-\langle y_\mathrm{m}\partial_y v_x^{(1)}\rangle\label{v2bc}
\end{equation}
and
\begin{eqnarray}
&&\langle-\partial_y\theta^{(2)}+w\theta^{(2)}\rangle_{y=0}\nonumber\\
&=&\langle -\partial_xy_\mathrm{m}\partial_x\theta^{(1)}+y_\mathrm{m}\partial_y^2\theta^{(1)}-wy_\mathrm{m}\partial_y\theta^{(1)}\rangle_{y=0}.\label{theta2bc}
\end{eqnarray}
The terms on the right-hand sides of~(\ref{v2bc}--\ref{theta2bc}) arise from expanding the boundary about $y=0$, and depend on the anchoring strength $w$ implicitly through the first-order quantities. However, for large Er these quantities are given by the strong-anchoring limit; in particular, all the terms on the right-hand side of ~(\ref{theta2bc}) vanish since $y_\mathrm{m}$ and $\theta^{(1)}$ are out of phase for strong anchoring. Furthermore, the value of $\partial_y\theta^{(2)}$ at $y=0$ is $\mathcal{O}(\sqrt{\mathrm{Er}})$. Since the angle field $\theta^{(2)}$ is $\mathcal{O}(\mathrm{Er})$ [see Eq.~(\ref{strongtheta})], the term $\langle\partial_y\theta^{(2)}\rangle_{y=0}$ is subleading in Er. Therefore, large Er behavior of the arbitrary anchoring strength problem is the strong-anchoring problem. The steady-state angle field evolves slowly, with a characteristic time proportional to Er.
}

\subsection{Asymptotic solution for low Ericksen number}
\label{sec:8}

\begin{figure*}[t]
\begin{center}
\includegraphics[height=1.4in]{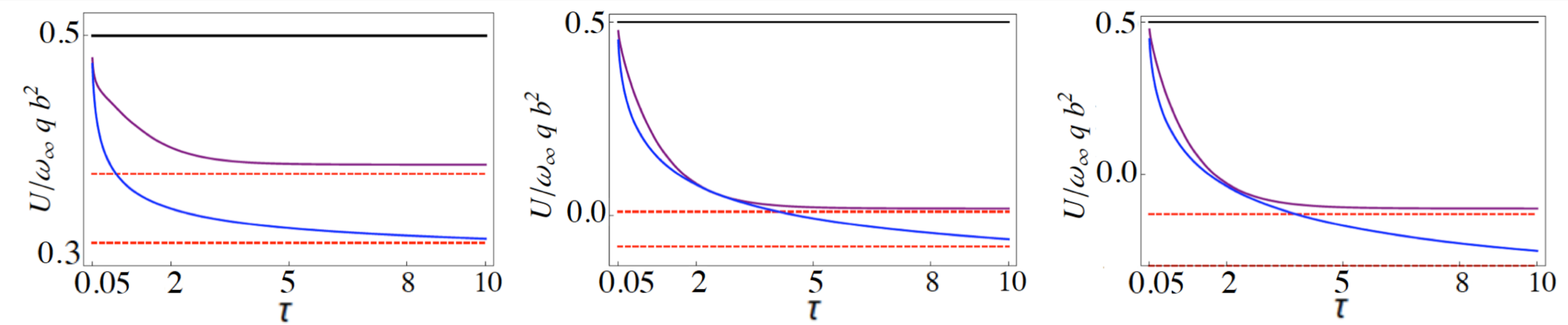}
\caption{(Color online)
Plot\trp{s} of \trp{the} dimensionless swimming speed $U/(\omega\trp{_\infty} q b^2)$ vs. $\tau$ for $\mathrm{Er} \ll1$ and $\log(\gamma / \mu) = 0$ (left panel), $\log(\gamma / \mu) = 1$ (middle panel), and $\log(\gamma / \mu) = 3$ (right panel). Colors correspond to $w=0.001$ (black), $w=1$ (purple), and $w=1000$ (blue). Red dashed lines indicate the steady swimming speed found in~\cite{KriegerSpagnoliePowers2014}.}
\label{smallErsols}
\end{center}
\end{figure*}

\trp{In this section we show explicitly for arbitrary dimensionless anchoring strength $w$ that the angle field attains its steady-state value quickly when the Ericksen number is low, with a startup time proportional to Er.}
Consider the %expansion of 
\trp{governing equations}%Eq\trp{s}.
~(\ref{fbal}) \trp{and} (\ref{tbal}) %, \ref{psieqn}, \ref{thetaeqn}) 
%for
\trp{in the limit of} $\trp{\mathrm{Er}\ll}1$. In this limit the viscous stresses are weak compared to elastic stresses. 
%At each order in amplitude, we expand in Ericksen number, denoting the power of Er by a subscript. For example, $\psi^{(1)}=\psi^{(1)}_0+\mathrm{Er}\psi^{(1)}_1+....$ 
\trp{Note that sending Er to zero removes the highest (and only) time derivative in Eq.~(\ref{tbal}). Therefore, we have a singular perturbation theory problem, with an inner problem for early time and an outer problem for late time. The outer problem is the steady-state problem studied in~\cite{KriegerSpagnoliePowers2014}.}
\trp{We introduce the short time scale $\tau=t\,\mathrm{Er}$ to resolve the fast relaxation of the angle field and flow to their steady-state values. Thus, we formally expand the fields at each order of $\epsilon$ in powers of Er.} For example, $\psi^{(1)}\trp{(t,\mathrm{Er})}=\psi^{(1)}_0\trp{(t,\tau)}+\mathrm{Er}\psi^{(1)}_1\trp{(t,\tau)}+....$ 
\trp{Note that 
\begin{equation}
\partial_t\theta^{(1)}=\frac{1}{\mathrm{Er}}\partial_\tau\theta^{(1)}_0+\partial_t\theta^{(1)}_0+\partial_\tau\theta^{(1)}_1+....
\end{equation}
Only the angle field enters the zeroth order equations:}
%We note that to solve this problem as a regular perturbation removes the highest time derivative. Solutions to this outer problem can be found in \cite{KriegerSpagnoliePowers2014}. Since the outer solution does not smoothly converge to the Taylor solution at $t=0$, we introduce an inner problem with a "fast" timescale $\tau \equiv \frac{t}{Er} = T$, equivalent to the beat timescale of the swimmer. Therefore $\partial _t \theta = \frac{1}{Er} \partial_{\tau} \theta$ and there is a boundary layer of thickness $Er$ near $t=0$. To zeroth order in $Er$ and first order in $\epsilon$ the system becomes
\begin{eqnarray}
\nabla ^4 \theta_0 ^{(1)} &=& 0  \\
\partial _{\tau} \theta_0 ^{(1)} &=& \frac{\mu}{\gamma} \nabla ^2 \theta_{0}^{(1)}.
\end{eqnarray} 
\trp{Expanding the boundary condition in Er leads to}
\begin{equation}
 w(\theta ^{(1)}_0 |_{y=0} - \epsilon \cos(x)) = \partial _y \theta^{(1)}_0 |_{y=0}.
\end{equation}
\trp{To zeroth order in Er, the solution is}
\begin{equation}
\theta_0 ^{(1)} = \epsilon \frac{w}{1+w} e^{-y} \cos x.
\end{equation}
\trp{Note that since the wave of the swimmer is stationary on the fast time scale $\tau$, the angle field is simply the equilibrium angle field for a stationary ripple.}

%To first order in $Er$, 
\trp{Turning now to the next order in Er,} we have
\begin{eqnarray}
\nabla ^4 \psi _0 ^{(1)} + \frac{1}{2} \nabla ^4 \theta_1 ^{(1)} &=& 0\label{lowEr1}  \\
\partial _{\tau} \theta _1 ^{(1)} + \frac{1}{2} \nabla ^2 \psi _0 ^{(1)} &=& \frac{\mu}{\gamma} \nabla ^2 \theta _1 ^{(1)},\label{lowEr2}
\end{eqnarray}
with boundary conditions
\begin{eqnarray}
(\partial\trp{_y} \psi^{(1)}_{0},-\partial_\trp{x} \psi^{(1)}_{0})_{y=0} - (0,-\epsilon \cos x)&=&0\label{lowEr3}  \\
 \left[w(\theta ^{(1)}_1 - \epsilon \tau \sin x)- \partial _y \theta^{(1)}_1 \right]_{y=0}&=&0\label{lowEr4}
\end{eqnarray}
\trp{and initial conditions $\psi^{(1)}_0=0$ and $\theta^{(1)}_1=0$. The form of Eqs.~(\ref{lowEr1}--\ref{lowEr2}) is similar to the form of (\ref{laplin1}--\ref{laplin2}), and the Laplace transforms of $\psi^{(1)}_0$ and $\theta^{(1)}_1$ are readily found. However, we are not able to find an explicitly analytic expression for the inverse Laplace transform of these quantities. The boundary conditions~(\ref{lowEr3}--\ref{lowEr4}) and the governing equations~(\ref{lowEr1}--\ref{lowEr2}) imply $\psi_0^{(1)}\propto\sin x$ and $\theta_1^{(1)}\propto\sin x$. These phase relations apply because the swimmer wave is stationary on the small timescale $t\,\mathrm{Er}$ over which the fields develop from their initial values.}

%which has the same general form as Eq.~(\ref{psieqn}), Eq.~(\ref{thetaeqn}) with modified boundary conditions, and must therefore be solved in Laplace space.  
\trp{Now consider the problem to second order in $\epsilon$.}
To zeroth order in Er, the second-order problem \trp{is} %becomes
\begin{eqnarray}
\partial _y ^3 \langle \theta _0 ^{(2)}\rangle &=& 0 \nonumber \\
\partial _{\tau} \theta_0 ^{(2)} &=& \frac{\mu}{\gamma} \partial _y ^2 \langle \theta^{(2)}_0\rangle,
\end{eqnarray}
%This suggests 
suggest\trp{ing that} $\langle \theta_0^{(2)} \rangle$ is a constant. The boundary condition at this order is 
\begin{equation}
\big[- \langle \partial _y \theta_0 ^{(2)} \rangle + w \langle \theta_0 ^{(2)} \rangle \big] _{y=0} = 0,
\end{equation}
\trp{implying}
%so that 
$ \langle \theta_0 ^{(2)} \rangle = 0$.  

To first order in Er, we have
\begin{eqnarray}
\partial _y ^2 \langle v_{x0}^{(2)}\rangle + \frac{1}{2} \partial _y ^3 \langle \theta_1 ^{(2)}\rangle &=& \langle  \partial _x \theta_0 ^{(1)} \nabla ^2 \theta _1 ^{(1)}\rangle \nonumber \\
\partial _{\tau} \langle  \theta_1^{(2)}\rangle + \frac{1}{2} \partial_y \langle  v_{x0}^{(2)}\rangle &=& \frac{\mu}{\gamma} \partial _y ^2 \langle \theta_1 ^{(2)}\rangle - \langle v_0 ^{(1)} \cdot \nabla \theta _0 ^{(1)}\rangle. \label{smallERfinal}
\end{eqnarray}
The boundary condition\trp{s to second order in $\epsilon$ and first order in Er are greatly simplified by the observation that $\theta^{(1)}_1$ and $y_\mathrm{m}$ are in phase in $x$, which makes many terms vanish when we expand $y_\mathrm{m}$ to first order in Er, $y_\mathrm{m}=\epsilon\sin x-\epsilon\mathrm{Er}\tau\cos x$. Thus}
\begin{eqnarray}
\trp{\langle v_{x0}^{(2)}\rangle|_{y=0}}&\trp{=}&\trp{-\epsilon\langle \partial_y v^{(1)}_{x0}\sin x\rangle}\\
\trp{\big[- \langle \partial _y \theta_1 ^{(2)} \rangle + w \langle \theta_1 ^{(2)} \rangle \big] _{y=0}}&\trp{=}&\trp{0}\label{lowEr2ndorder2}
%\langle (w-\partial_y) \theta^{(2)}_1\rangle|_{y=0} &=&  \langle \tau\partial_x \theta_1 ^{(1)}\sin x- \tau  \partial _y ^2 \theta_1 ^{(1)}\cos x \nonumber \\
%&-& w \tau cos(x) \partial _y \theta^{(1)} _1\rangle - \frac{w^2}{2(1+w)} sin(\tau)
\end{eqnarray}
%Because of the presence of the $\partial _{\tau} \langle \theta _1 ^{(2)} \rangle$ term, we cannot simply diagonalize Eq.~(\ref{smallERfinal}) and extract the swimming speed as in [\ref{ksp14}]. Instead we
%
%We can rearrange Eq.~(\ref{smallERfinal}) into a single equation for $\langle \theta^{(2)} _1 \rangle$: 
%\begin{equation}
%\partial_{\tau} \langle \theta_1 ^{(2)}\rangle = D \partial _y ^2 \langle \theta_1 ^{(2)}\rangle - g_0 - \frac{1}{2} \int f_1(y) dy \label{smallerQ}
%\end{equation}
%To %plot 
\trp{Since $\theta^{(1)}_0$ is independent of time, it is straightforward to analytically determine the Laplace transform Eq.~(\ref{smallERfinal}). Using the boundary condition~(\ref{lowEr2ndorder2}) to determine the Laplace transform of $\langle\theta^{(2)}_1\rangle$, we may use} the same methods as in Section~\ref{sec:6} to  solve for the swimming speed $\tilde{U}(s)$. 
%the swimming speed, we take the Laplace transform of  Eq.~(\ref{smallERfinal}) and its boundary condition \trp{(\ref{lowEr2ndorder2}). } and solve for the swimming speed $\tilde{U}(s)$ using the same methods as in Section 3. 
%This function does not have an 
\trp{We could not find an} exact inverse Laplace transform \trp{for this function} except in the limit $w \rightarrow \infty$, so we invert the expression numerically %to examine $U(t)$. For the numerical inversion we use 
\trp{using} a Fourier/de Hoog type method~\cite{daviesmartin} on the interval $t/\tau \in [0.05,20]$ with 2,048 gridpoints. %Solutions are plotted in Fig.~(\ref{smallErsols}).
\trp{Figure~\ref{smallErsols} shows that the swimming speed relaxes to the steady state value with a characteristic time proportional to Er for various anchoring strengths.}

\subsection{Zero anchoring}
\label{sec:10}

%\begin{figure*}[t]
%\begin{center}
%\includegraphics[height=1.5in]{Unumfig1}
%\caption{(Color online)
%Plot of dimensionless swimming speed $U/(\omega q b^2)$ vs. $t/t_e$ for $w=0$, $Er = 1$ and $log(\gamma / \mu) = 0$ (left panel), $log(\gamma / \mu) = 1$ (middle panel), and $log(\gamma / \mu) = 3$ (right panel). Solid lines correspond to the steady-state speeds computed in \cite{KriegerSpagnoliePowers2014}.}
%\label{Unumfig1}
%\end{center}
%\end{figure*}

\begin{figure}[t]
\begin{center}
\includegraphics[height=2.2in]{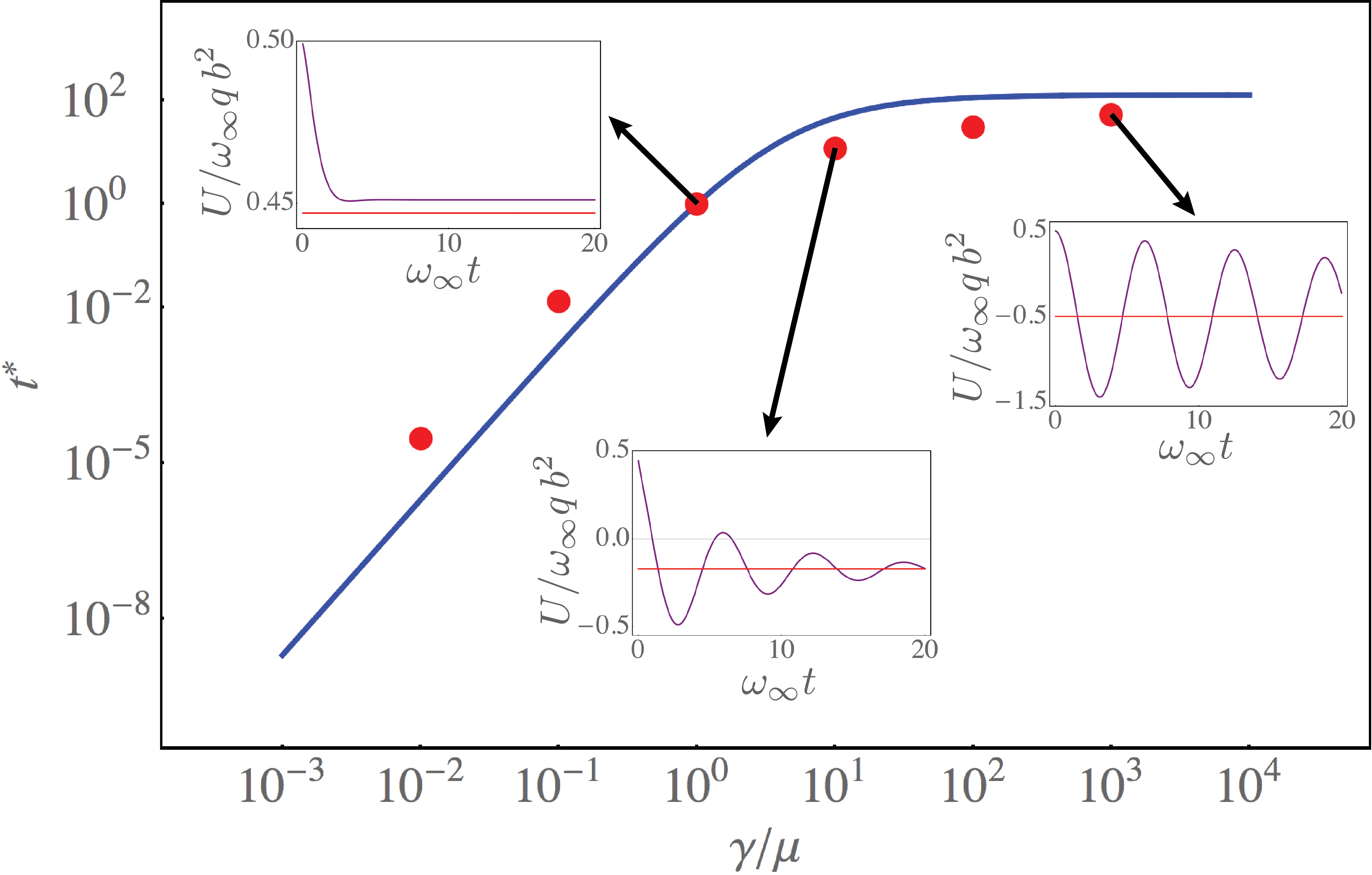}
\caption{(Color online) 
%Log-log plot of the settling time $t^*$ versus $\gamma / \mu$ for $Er = 1$. Here we take $t^*$ to be the earliest time satisfying the two conditions that $\frac{U^{\infty} - U(t^*)}{U^{\infty}} < 0.05$ and $U^{\infty} - U(t^*)$ is monotone \trp{monotonic?} $\forall t>t^*$.The solid blue line corresponds to the case of strong anchoring, and the red dots to numerical solutions for weak anchoring. $t^*$ is rescaled in such a way that $t^* = 1$ for $\gamma = \mu$. 
Log-log plot of the \trp{dimensionless time} $t^*$ {to attain the steady-state swimming speed} versus $\gamma / \mu$ for $\mathrm{Er} = 1$.  \trp{The time $t^*$ is defined to be the time in which the amplitude of the envelop of the oscillations (if there are any) is 5\% of the ultimate swimming speed.  The red dots are the values of $t^*$ for zero anchoring strength, $w=0$.  For comparison,} the solid blue line corresponds to the case of strong anchoring. \trp{To best compare the dependence of $\gamma/\mu$, the time-origins of both graphs have been shifted by defining $t^*=1$ for $\gamma=\mu$. The insets show the numerically computed swimming speeds versus dimensionless time for $\gamma/\mu=1$, $10$, and $1000$.}
}
\label{settimes}
\end{center}
\end{figure}

%The general second-order problem for arbitrary Ericksen number cannot be solved analytically due to to the nonlinear convective terms $f$, $g$, which have no analytical Laplace or inverse-Laplace transform. 

\trp{The final limit we consider is zero anchoring strength, $w=0$, for arbitrary Ericksen number. We proceed as above, by solving the governing equations to first order in $\epsilon$ analytically using the Laplace transform, finding the angle field and swimming velocity in the frequency domain to second order in $\epsilon$, and then using the Fourier/de Hoog method to invert the Laplace transform to find the swimming speed $U(t)$.  The results are shown in Fig.~\ref{settimes} for Er=1. When $\gamma/\mu$ is sufficiently small, less than around 10, the ultimate swimming direction is opposite the direction of the traveling waves of the stroke~\cite{KriegerSpagnoliePowers2014}. When $\gamma/\mu$ is large enough, the swimmer ultimately swims in the same direction as the traveling waves. Figure~\ref{settimes} shows that in these cases the swimming speed oscillates about the final swimming speed as the steady state develops; when $\gamma/\mu$ is large there can be several time intervals in which the swimmer changes direction before it settles down into its steady swimming speed.  }

\section{Conclusion}
\label{conclude}
\trp{In conclusion, we have studied the time-evolution of a swimmer  in a hexatic liquid crystal film. The swimmer in our approximation  has  a small amplitude sinusoidal stroke that begins abruptly. Since the startup of viscous flow is much faster than the evolution of the liquid crystal configuration, the swimmer immediately begins swimming with the Newtonian swimming speed. It reaches the ultimate steady-state swimming speed with a dimensionless characteristic time that is proportional to the Ericksen number, or, equivalently, a characteristic time proportional to $\mu/(K q^2)$. When the Ericksen number is large, the behavior is independent of that anchoring strength and given by the limit of infinite anchoring. When the anchoring strength vanishes, the transient swimming speed can oscillate, and even change sign. Although the high symmetry of a hexatic liquid crystal lead to great simplifications in our analysis, we expect that many of the phenomena we found here will serve as a guide for the investigation of more the more realistic but complicated case of a finite-size swimmer in a three-dimensional nematic liquid crystal.}

\begin{acknowledgements}
This work was supported in part by National Science Foundation Grant Nos.  CBET-1437195 (TRP) and CBET-1336638 (TRP).
\end{acknowledgements}

% BibTeX users please use
\bibliographystyle{unsrt}
 %\bibliography{newrefs}
\bibliography{newrefs} %tom will move this bib file into this directory later

%%
%% For one-column wide figures use
%\begin{figure}
%% Use the relevant command for your figure-insertion program
%% to insert the figure file.
%% For example, with the option graphics use
%%\resizebox{0.75\textwidth}{!}{%
%  %\includegraphics{leer.eps}
%%}
%% If not, use
%%\vspace{5cm}       % Give the correct figure height in cm
%\end{figure}
%%
%% For two-column wide figures use
%\begin{figure*}
%% Use the relevant command for your figure-insertion program
%% to insert the figure file. See example above.
%% If not, use
%\vspace*{5cm}       % Give the correct figure height in cm
%\end{figure*}
%%\begin{table}
%%\caption{Please write your table caption here}
%%\label{tab:1}       % Give a unique label
%%\begin{tabular}{lll}
%%\hline\noalign{\smallskip}
%%first & second & third  \\
%%\noalign{\smallskip}\hline\noalign{\smallskip}
%%number & number & number \\
%%number & number & number \\
%%\noalign{\smallskip}\hline
%%\end{tabular}
%% Or use
%\vspace*{5cm}  % with the correct table height
%%\end{table}
%%
%%% Non-BibTeX users please use
%\begin{thebibliography}{}
%%
%% and use \bibitem to create references.
%Author, \textit{Book title} (Publisher, place year) page numbers
%\end{thebibliography}

\end{document}